\pdfoutput=1
\documentclass[runningheads]{llncs}
\usepackage[T1]{fontenc}
\usepackage{tabularray}

\usepackage{bbm,amsbsy,amsmath,amsfonts, amssymb,bm}
\usepackage{rotate, array, color,colordvi, psfrag}
\usepackage{xcolor}
\usepackage{cite}
\usepackage{booktabs}
\usepackage{multirow}

\usepackage{hyperref}
\usepackage{graphicx}

\begin{document}
\title{NeuroSymAD: A Neuro-Symbolic Framework for Interpretable Alzheimer's Disease Diagnosis}
\author{
Yexiao He\inst{1}$^\dag$ \and
Ziyao Wang\inst{1}$^\dag$ \and
Yuning Zhang\inst{1} \and
Tingting Dan\inst{2} \\ 
Tianlong Chen\inst{2} \and
Guorong Wu\inst{2} \and
Ang Li\inst{1}}
\authorrunning{Yexiao He et al.}
%
\institute{University of Maryland, College Park, MD 20742, USA \\
\email{\{yexiaohe, ziyaow, yuning, angliece\}@umd.edu} 
\and
University of North Carolina at Chapel Hill, Chapel Hill, NC 27599, USA \\
\email{tianlong@cs.unc.edu}\\
\email{\{tingting\_dan, guorong\_wu\}@med.unc.edu}
}


%
\titlerunning{NeuroSymAD: Neuro-Symbolic Framework for Alzheimer's Disease Diagnosis}

\maketitle              
\let\thefootnote\relax
\footnotetext{$\dag$ Equal Contribution.}

\begin{abstract}

Alzheimer's disease (AD) diagnosis is complex, requiring the integration of imaging and clinical data for accurate assessment. While deep learning has shown promise in brain MRI analysis, it often functions as a black box, limiting interpretability and lacking mechanisms to effectively integrate critical clinical data such as biomarkers, medical history, and demographic information.
To bridge this gap, we propose NeuroSymAD, a neuro-symbolic framework that synergizes neural networks with symbolic reasoning. A neural network percepts brain MRI scans, while a large language model (LLM) distills medical rules to guide a symbolic system in reasoning over biomarkers and medical history. This structured integration enhances both diagnostic accuracy and explainability.
Experiments on the ADNI dataset demonstrate that NeuroSymAD outperforms state-of-the-art methods by up to 2.91\% in accuracy and 3.43\% in F1-score while providing transparent and interpretable diagnosis.

\end{abstract}

\section{Introduction}
Alzheimer's disease (AD)~\cite{blennow2006alzheimer, scheltens2016alzheimer} is one of the most pressing healthcare challenges in the aging society, affecting millions globally—a number projected to triple by 2050. This surge not only threatens public health but also imposes profound socioeconomic burdens.
Accurate diagnosis of AD is crucial for timely intervention and better patient outcomes~\cite{deture2019neuropathological}. Currently, AD diagnosis relies heavily on the comprehensive analysis of multiple data modalities, including brain MRI scans~\cite{jack2008alzheimer}, various biomarkers~\cite{jack2013biomarker}, and patient history~\cite{bateman2012clinical}. Clinicians follow a sophisticated diagnostic process that integrates visual image analysis with reasoning based on extensive medical knowledge and expertise. This process exemplifies a unique integration of perceptual skills in image analysis and logical reasoning with domain expertise, effective yet challenging to automate.

Recent deep learning advances in brain MRI analysis for AD diagnosis have achieved great success~\cite{tanveer2020machine, mirzaei2016imaging}. However, they face several critical limitations. First, they typically operate as isolated components focusing solely on imaging features, failing to integrate other crucial clinical information such as biomarkers and patient history. 
Second, their black-box nature poses challenges for clinical adoption, as healthcare professionals require transparent and interpretable decision-making. Third, despite the abundant medical knowledge accumulated in clinical guidelines, research papers, and expert experience, current deep learning systems lack effective mechanisms to incorporate existing valuable domain expertise, not to mention discovering new knowledge. In contrast, rule-based symbolic systems can explicitly encode medical knowledge but struggle with handling visual data and require substantial manual effort in rule construction and updating. 
These limitations motivate the development of a neuro-symbolic system that merges deep learning’s perceptual strengths with the interpretability of symbolic reasoning, mimicking clinicians’ diagnostic process.

In this paper, we propose NeuroSymAD, a novel neuro-symbolic AD diagnosis framework that seamlessly integrates deep learning-based image perception with knowledge-driven symbolic reasoning. 
Our contributions are: (1) We are the first neuro-symbolic framework for AD diagnosis that effectively mimics clinical experts' diagnostic process by combining neural network perception with symbolic reasoning. (2) We design an automated knowledge acquisition module to automatically construct and update the symbolic reasoning system, improving the efficiency and scalability of medical knowledge integration compared to manual construction. (3) We use an end-to-end training strategy that jointly optimizes the neural and symbolic components, allowing medical knowledge to guide neural network learning while simultaneously adapting symbolic rules based on empirical data. (4) Extensive experiments on clinical datasets that demonstrate superior diagnostic accuracy and interpretability.

\section{Related Literature}

Recent advances in deep learning have proven effective for AD diagnosis using neuroimaging. 3D CNNs, and 3D ResNets have achieved high accuracy in classifying AD from MRI scans \cite{mirzaei2016imaging, alsubaie2024alzheimer, abdulazeem2021cnn, al2022alzheimer, turrisi2023effectdataaugmentation3dcnn, ebrahimi2020introducing, farooq2017deep, mohi2023novel}. However, such models typically operate as “black boxes” and are limited by single-modality inputs. 
To address this, multimodal frameworks integrating imaging, genetic, and clinical data have been developed\cite{qiu2022multimodal,golovanevsky2022multimodal, liu2018use, elazab2024alzheimer, zhanga2023multimodalgraphneuralnetwork, venkatraman2024dual}. 
Some works have innovated on model architectures to enhance performance \cite{zhanga2023multimodalgraphneuralnetwork, venkatraman2024dual, alsubaie2024convadd, choudhury2024coupled, yaqoob2024prediction, shaffi2024ensemble}. 
Neuro-symbolic method has also been proposed. The PP-DKL \cite{lavin2021neurosymbolicneurodegenerativediseasemodeling} uses probabilistic programming to predict early neurodegeneration. 
Our method offers several distinct advantages over the existing methods. Unlike PP-DKL \cite{lavin2021neurosymbolicneurodegenerativediseasemodeling}, which relies on single-modality data, NeuroSymAD integrates MRI, demographic, and clinical data for a robust diagnosis. It also improves interpretability and scalability by incorporating a symbolic reasoning module that leverages automated knowledge extraction. Finally, its end-to-end optimization enables effective collaboration between neural and symbolic components, further improving diagnostic accuracy.

\section{Methods}
\begin{figure}[t]
  \centering
  \includegraphics[,width=0.7\textwidth]{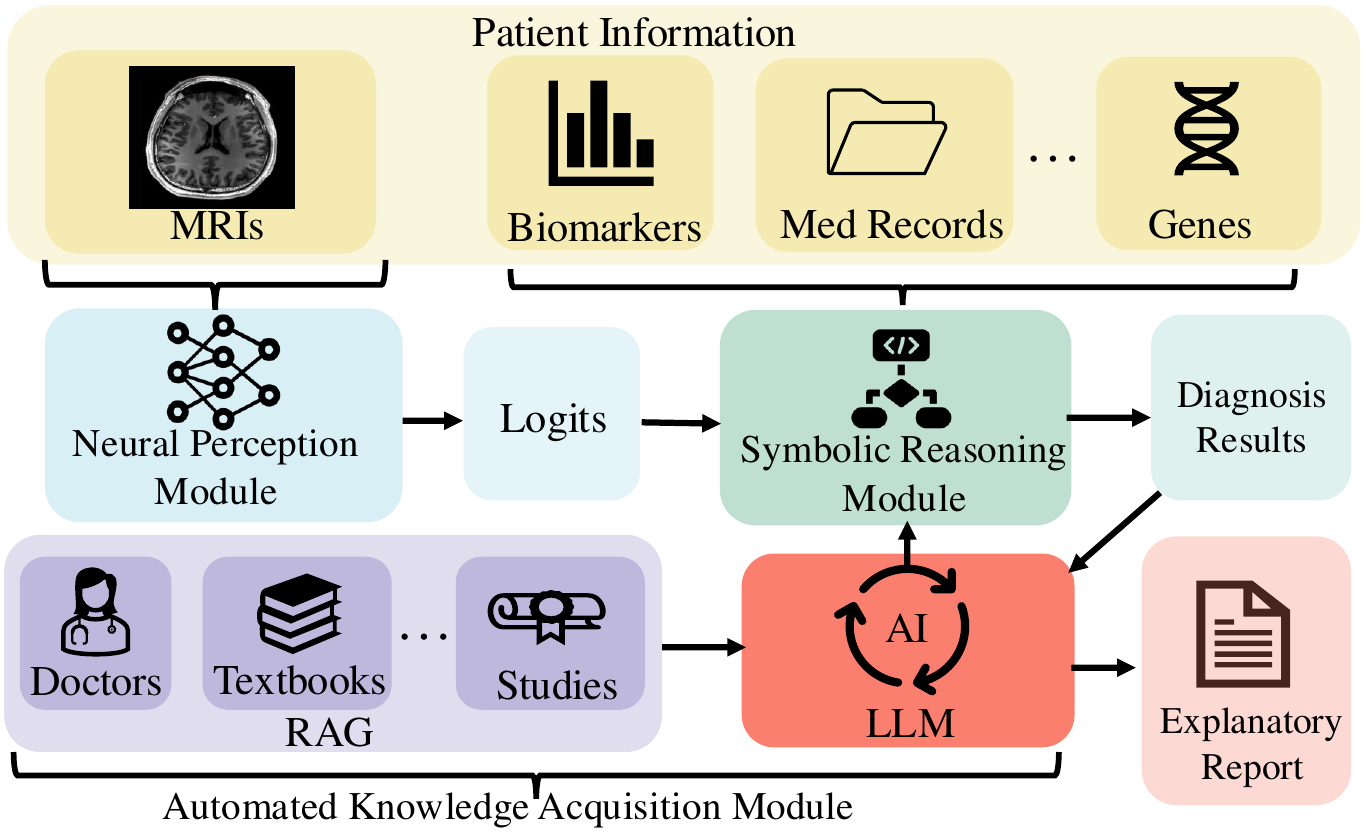} 
  \caption{Overview of NeuroSymAD} 
  \label{fig:framework}  
\end{figure}
Our AD diagnosis neuro-symbolic framework comprises three key components: a neural perception module, a symbolic reasoning module, and an automated knowledge acquisition module, as illustrated in Figure~\ref{fig:framework}. Given a patient sample—including a 3D MRI scan $\mathbf{x}_i \in \mathbb{R}^{m \times n \times r}$ and a set of patient's information and medical history records $\mathbf{z}_i \in \mathbb{R}^{k}$—the neural perception module employs a deep neural network to process the MRI scan and estimate the probability of AD. This estimation is represented as logits $\mathbf{y_i} \in \mathbb{R}^{2}$, corresponding to two classes: cognitively normal (CN) and AD. 
Next, the symbolic reasoning module refines these logits based on the patient information $\mathbf{z}_i$ by leveraging a set of learned rules to enhance diagnostic accuracy. These rules are generated by the automated knowledge acquisition module, which is composed of a Large Language Model (LLM) integrated with a Retrieval-Augmented Generation (RAG) system. Furthermore, 
an end-to-end training strategy is employed to jointly optimize the neural perception module and symbolic reasoning module. 
Finally, the automated knowledge acquisition module consolidates the adjusted logits and reasoning rules using the LLM to generate an explanatory report to show the reasoning process, supporting clinical decision-making.

\subsection{Neural Perception Module}
The neural perception module processes MRI scans using a deep learning model, denoted as $\mathcal{F}$, mapping an MRI scan to probability logits:

\begin{equation}
\begin{split}
    \mathcal{F}: \mathbb{R}^{m \times n \times r} \rightarrow \mathbb{R}^{2}, \quad
    \mathbf{y_i} = \mathcal{F}(\mathbf{x_i}).
    \label{eq:nn}
\end{split}
\end{equation}
where $\mathbf{y_i}$ represents the predicted probability logits for the two classes.
To optimize the model, we first pretrain it solely on MRI scans using a cross-entropy loss function, optimized via an Adam optimizer \cite{kingma2017adammethodstochasticoptimization}:

\begin{equation}
\mathcal{L} = -\frac{1}{N} \sum_{i=1}^{N} \left[ \hat{y}_i \log y_i + (1 - \hat{y}_i) \log (1 - y_i) \right],
\label{eq:loss}
\end{equation}
where \(\hat{y}_i\) is the ground-truth diagnosis label for \(\mathbf{x}_i\), and \(y_i\), obtained as \(y_i = \operatorname{softmax}(\mathbf{y}_i)\), is the predicted label. After this MRI-only pretraining, the model is further refined through end-to-end training with the symbolic reasoning module.


\subsection{Symbolic Reasoning Module}

The symbolic reasoning module enhances the neural network's predictions by integrating domain knowledge in the form of symbolic rules. An LLM equipped with an AD clinical knowledge RAG system generates these rules, capturing the relationships between demographic, clinical characteristics, medical history records, and classification logits.
Given the initial logits $\mathbf{y}_i$ from the neural network, we apply a set of symbolic rules $\mathcal{R} = \{R_1, R_2, ..., R_K\}$ generated by the LLM. The symbolic reasoning process first analyzes the patient information, biomarkers and medical history records $\mathbf{z}_i$ to determine which subset of rules $\mathcal{R}_i \subseteq \mathcal{R}$ are relevant for the current patient. Then, these identified rules are applied through a set of differentiable operations to adjust the logits.

For each rule, we define specific differentiable operations with learnable parameters that capture various medical relationships. The age-related rule provides an example of how medical knowledge is encoded in our framework:

\newcommand{\relu}{\operatorname{ReLU}}

\begin{equation}
\delta_{age} = \alpha \cdot \sigma\left(\frac{z_{age} - T_{1}}{\tau}\right) + \beta \cdot \relu(z_{age} - T_{2})
\end{equation}
where $\alpha, \beta, T_1, T_2$ are trainable parameters. $\alpha$ is the base effect strength, $T_{1}$ is the age threshold, $\beta$ is the acceleration factor, and $T_{2}$ is the acceleration threshold. $\tau$ controls the smoothness of the sigmoid transition $\sigma(\cdot)$, which models a smooth transition in risk at the threshold age, while the second term captures accelerated risk increase in advanced age. This formulation encodes the clinical knowledge that AD risk increases after a certain age and accelerates further in later years \cite{isik2010late}. By training these parameters, the model automatically determines at what ages these effects become significant and how strong each effect is.

Similar differentiable formulations are defined for other rules, including gender-specific factors, education level, comorbidities, lifestyle factors, clinical indicators, etc. $\delta_i = \sum_{j \in \mathcal{R}_i} \delta_{i,j}$ aggregates the effects of all relevant rules.
The final adjusted logits are obtained by applying this cumulative adjustment:

\begin{equation}
\tilde{\mathbf{y}}_i = \mathbf{y}_i + [ -w \cdot \delta_i, \ \delta_i ]^T
\end{equation}
where $w$ is a balance factor. The flexible formulations allow us to incorporate complex domain-specific relationships in a differentiable manner, enabling end-to-end training. The learnable parameters of each rule operation are optimized during training, allowing the model to automatically determine the appropriate effect of each medical factor based on empirical data while maintaining the structure informed by clinical knowledge.

After pretraining the neural perception module on MRIs alone, we perform end-to-end training by incorporating patients' information and medical history. In this phase, both the neural network weights and the trainable rule parameters in the symbolic system are jointly optimized. This not only allows the symbolic rules to align better with empirical data but also leverages medical knowledge to guide the optimization of the neural network. This bidirectional interaction between neural and symbolic components enables NeuroSymAD to achieve high accuracy and interpretability, making it more suitable for clinical applications.

\subsection{Automated Knowledge Acquisition Module}
The knowledge acquisition module performs automatic rule generation and produces explanatory reports. For rule generation, it leverages LLM with RAG to extract symbolic rules from clinical guidelines, research papers, and textbooks. The process involves retrieving relevant documents, extracting clinical insights, and transforming them into formal logical statements that capture relationships between biomarkers, medical history, and diagnostic criteria. These rules are then converted into a differentiable format with trainable parameters. This approach significantly reduces manual effort while maintaining comprehensive and up-to-date rulesets.
During inference, this module generates explanatory reports, which include the diagnostic result and reasoning behind the diagnosis.

\section{Experiments}
\noindent \textbf{\emph{Dataset:}} We evaluated NeuroSymAD using the Alzheimer's Disease Neuroimaging Initiative (ADNI1-3) dataset \cite{mueller2005alzheimer}. This dataset includes T1-weighted MRI scans from 3088 individuals. Skull stripping was performed using FreeSurfer's recon-all tool \cite{fischl2012freesurfer} to remove non-brain tissues. Each subject is associated with clinical diagnostic labels that span different cognitive states, including Cognitive Normal (CN), Subjective Memory Complaints (SMC), Early Mild Cognitive Impairment (EMCI), Late Mild Cognitive Impairment (LMCI), and Alzheimer’s Disease (AD). To address data imbalance issues, we consolidated these categories into two broader groups based on disease severity. Specifically, we merged CN, SMC, and EMCI into a single "CN" group, representing individuals with less severe conditions, while LMCI and AD are combined into an "AD" group, representing more advanced disease stages. Additionally, the dataset provides various demographic and clinical characteristics for each subject.
\begin{table*}[t]
\centering
\caption{Performance comparison of NeuroSymAD against SOTA deep learning methods for AD diagnosis (mean $\pm$ standard deviation over 10 random seeds). Best results are highlighted in \textbf{bold}.}
\label{tab:exp1}
\resizebox{0.8\textwidth}{!}{
\begin{tabular}{c|ccccc} 
\toprule
\textbf{Method} & \textbf{Accuracy} & \textbf{Precision} & \textbf{Recall} & \textbf{F1} & \textbf{AUC} \\
\midrule
DenseNet-169                         & 81.64$_{\pm 1.84}$   & 82.40$_{\pm 0.81}$   & 79.21$_{\pm 1.32}$   & 80.78$_{\pm 0.78}$   & 81.54$_{\pm 1.32}$   \\
Opt 3D CNN                     & 83.30$_{\pm 1.52}$   & 80.21$_{\pm 2.02}$   & 92.42$_{\pm 1.78}$   & 85.88$_{\pm 0.89}$   & 89.20$_{\pm 1.24}$   \\
DA CNN                   & 86.42$_{\pm 1.32}$   & 86.88$_{\pm 1.24}$   & 93.24$_{\pm 1.45}$   & 88.46$_{\pm 0.54}$   & 89.21$_{\pm 0.87}$   \\
3D ResNet                   & 85.67$_{\pm 1.20}$   & 86.16$_{\pm 3.08}$   & \textbf{95.25}$_{\pm 2.37}$   & 90.46$_{\pm 0.64}$   & \textbf{93.36}$_{\pm 0.47}$   \\
\midrule
\textbf{NeuroSymAD} & \textbf{88.58}$_{\pm 1.75}$ & \textbf{89.97}$_{\pm 0.83}$ & 94.44$_{\pm 1.69}$ & \textbf{92.15}$_{\pm 0.93}$ & 92.56$_{\pm 1.59}$  \\
\bottomrule
\end{tabular}}
\end{table*}

\begin{table*}[h]
\centering
\caption{Comparison of our method with existing neural networks. Starred metrics (*) indicate a significant improvement ($p<0.05$, paired t-test) with symbolic components.}
\resizebox{0.86\textwidth}{!}{
\begin{tabular}{c|c|ccccc}
\toprule
\textbf{Model}  &\textbf{Method}    & \textbf{Accuracy}       & \textbf{Precision}      & \textbf{Recall}         & \textbf{F1}            & \textbf{AUC}           \\
\midrule
\multirow{2}{*}{DenseNet-169}  &Base   & 81.64$_{\pm 1.84}$      & 82.40$_{\pm 0.81}$      & 79.21$_{\pm 1.32}$      & 80.78$_{\pm 0.78}$      & 81.54$_{\pm 1.32}$      \\
&Ours  & \textbf{83.63$^*$$_{\pm 0.84}$}  & \textbf{85.24$^*$$_{\pm 1.24}$}  & \textbf{83.20$^*$$_{\pm 0.90}$}  & \textbf{84.21$^*$$_{\pm 0.98}$}  & \textbf{82.31$_{\pm 1.42}$}      \\
\midrule
\multirow{2}{*}{Opt 3D CNN} &Base  & 83.30$_{\pm 1.52}$      & 80.21$_{\pm 2.02}$      & 92.42$_{\pm 1.78}$      & 85.88$_{\pm 0.89}$      & 89.20$_{\pm 1.24}$      \\
&Ours    & \textbf{85.81$^*$$_{\pm 2.13}$}  & \textbf{83.45$^*$$_{\pm 1.86}$}  & \textbf{93.05$_{\pm 2.03}$}      & \textbf{87.98$^*$$_{\pm 1.92}$}  & \textbf{90.10$^*$$_{\pm 0.88}$}  \\
\midrule
\multirow{2}{*}{DA CNN}  &Base  & 86.42$_{\pm 1.32}$      & 86.88$_{\pm 1.24}$      & \textbf{93.24$_{\pm 1.45}$}      & 88.46$_{\pm 0.54}$      & 89.21$_{\pm 0.87}$      \\
&Ours   & \textbf{87.32$_{\pm 1.12}$}      & \textbf{88.35$^*$$_{\pm 2.32}$}  & 91.23$^*$$_{\pm 1.65}$  & \textbf{89.76$^*$$_{\pm 1.40}$}  & \textbf{89.40$_{\pm 0.35}$}      \\
\midrule
\multirow{2}{*}{3D ResNet}  &Base  & 85.67$_{\pm 1.20}$      & 86.16$_{\pm 3.08}$      & \textbf{95.25$_{\pm 2.37}$} & 90.46$_{\pm 0.64}$      & \textbf{93.36$_{\pm 0.47}$} \\
&Ours               & \textbf{88.58$^*$$_{\pm 1.75}$} & \textbf{89.97$^*$$_{\pm 0.83}$} & 94.44$_{\pm 1.69}$  & \textbf{92.15$^*$$_{\pm 0.93}$} & 92.56$_{\pm 1.59}$      \\
\bottomrule
\end{tabular}}
\label{tab:comparison}
\end{table*}

\noindent \textbf{\emph{Baselines:}} To evaluate NeuroSymAD, we compared it against four state-of-the-art (SOTA) deep learning approaches for AD diagnosis. The first is DenseNet-169, which leverages dense connectivity patterns to maximize information flow between layers, demonstrating strong performance in capturing hierarchical features from brain MRI scans \cite{al2022alzheimer}. The second is an optimized 3D CNN (Opt 3D CNN) developed specifically for AD diagnosis via MRI analysis \cite{turrisi2023effectdataaugmentation3dcnn}. The authors evaluated several architectures and found the best-performing one. The third is a Dual-Attention CNN (DA CNN) that incorporates spatial and channel attention mechanisms, enabling the focused analysis of relevant brain regions and feature channels for AD detection \cite{venkatraman2024dual}. The fourth is 3D ResNet, which adapts the residual learning framework to 3D medical imaging \cite{ebrahimi2020introducing}. These baselines represent different architectural paradigms in deep learning-based AD diagnosis, providing a comprehensive foundation for evaluating NeuroSymAD.

\noindent \textbf{\emph{Metrics:}} To evaluate NeuroSymAD, we employed standard metrics (accuracy, precision, recall, and F1-score). We also computed Area Under the Curve (AUC) to measure the model’s discriminative ability between AD and CN cases. 

\noindent \textbf{\emph{Implementation details:}} NeuroSymAD was implemented using PyTorch with a two-stage training strategy. The first stage focuses on training the neural perception module, allowing the backbone network to develop robust perceptual capabilities before incorporating domain knowledge. In the second stage, we incorporated the symbolic reasoning module and fine-tuned the entire system end-to-end with a lower learning rate. For our experiments, we processed MRI scans resampled to $128 \times 128 \times 128$ resolution. For data preprocessing, we applied standard normalization and data augmentation techniques including random rotations and flips. The training protocol employed a weighted cross-entropy loss to handle class imbalance. We used the Adam optimizer with initial learning rates of $10^{-4}$ and $10^{-5}$ for the first and second stages respectively, along with a step learning rate scheduler ($\gamma = 0.5$, step size = 10). The model was trained for 30 epochs in each stage with a batch size of 8.

\noindent \textbf{\emph{Results and Discussion:}}
Table \ref{tab:exp1} presents the performance comparison of the NeuroSymAD (using 3D ResNet as the backbone) against four SOTA approaches based on pure neural networks for AD diagnosis. The results demonstrate that NeuroSymAD consistently outperforms all baseline methods across key evaluation metrics. NeuroSymAD achieves an accuracy of 88.58\%, representing a significant improvement of 2.16\% over the best-performing baseline (Dual-Attention CNN at 86.42\%). The high precision (89.97\%) indicates our model's effectiveness in reducing false positive diagnoses, while the strong recall (94.44\%) demonstrates its capability to identify actual AD cases. Notably, NeuroSymAD achieves the highest F1-score (92.15\%), indicating superior overall performance. The relatively low standard deviations across metrics demonstrate the robustness of NeuroSymAD across different random initializations, further validating our hypothesis that integrating neural perception with symbolic reasoning effectively mimics the diagnostic process of experienced clinicians.

\begin{figure}[t]
  \centering
  \includegraphics[width=0.80\textwidth]{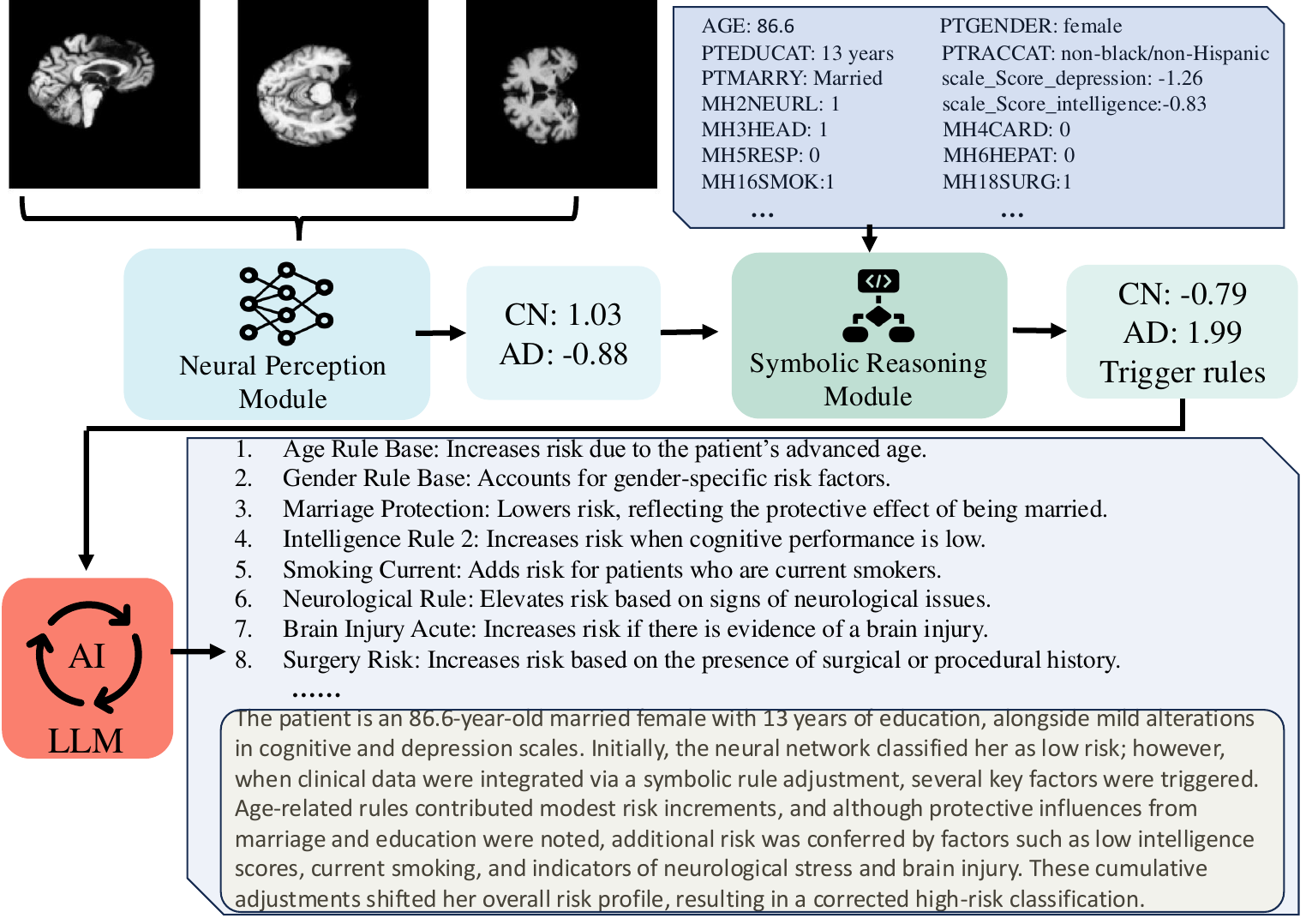} 
  \caption{An example of NeuroSymAD's diagnostic process and the generated explanatory report, demonstrating how the symbolic reasoning module corrects the misclassification of neural network.} 
  \label{fig:case}  
\end{figure}

Table~\ref{tab:comparison} compares the performance of different model architectures with and without our NeuroSymAD method. Across all models, the inclusion of symbolic reasoning module consistently improves performance. 3D ResNet achieves the best overall performance, with Accuracy increasing from 85.67\% to 88.58\%, Precision from 86.16\% to 89.97\%, and F1-score from 90.46\% to 92.15\% by adding the symbolic reasoning module. Similarly, Optimized 3D CNN and Dual-Attention CNN benefit from symbolic integration, showing significant gains in Accuracy, Precision, and F1-score.  
While Recall remains relatively stable, Precision improves across all models, reducing false positives. The AUC scores remain high, indicating robust model discrimination. A paired t-test confirms that improvements in key metrics are statistically significant.  
The results demonstrate that symbolic reasoning consistently enhances deep learning models.

Figure \ref{fig:case} shows an example of NeuroSymAD's diagnostic process and the generated explanatory report, illustrating a key advantage of NeuroSymAD. The case involves an 86.6-year-old female patient whose MRI scan was initially classified as CN (CN: 1.03, AD: -0.88) by the neural perception module alone. However, when patient information was processed through the symbolic reasoning module, which incorporates medical knowledge through rule-based analysis, the classification was correctly revised to indicate AD (CN: -0.79, AD: 1.99).
The symbolic reasoning module activated 15 different rules that adjusted the initial logits based on risk factors such as advanced age, neurological issues, smoking status, and cognitive performance indicators. This integration of clinical data with imaging analysis more accurately mimics the diagnostic process of experienced clinicians. The case demonstrates how NeuroSymAD overcomes the limitations of pure deep learning approaches by incorporating domain knowledge and patient-specific factors, providing both improved diagnostic accuracy and transparent reasoning through a comprehensive LLM-generated report explaining the factors that influenced the classification.

\begin{figure}[h]
  \centering
  \includegraphics[,width=\textwidth]{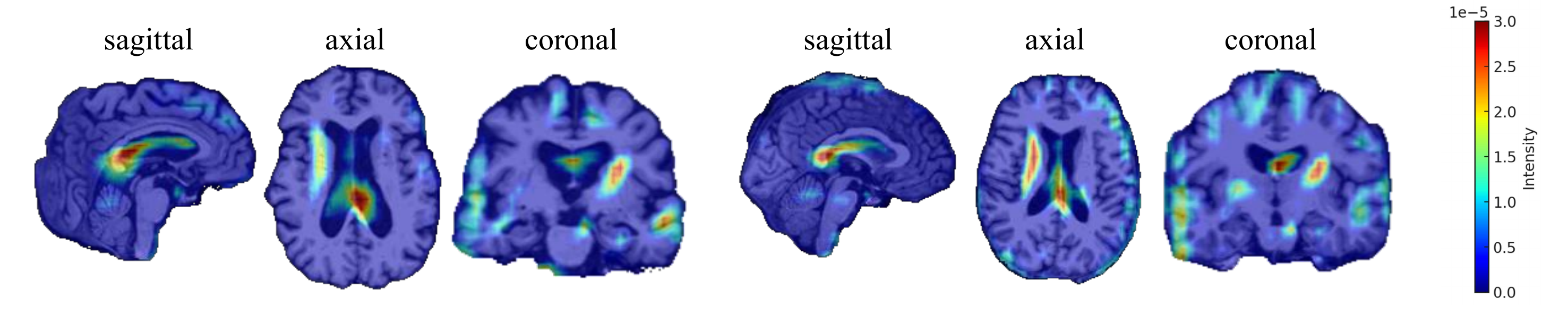} 
  \caption{Heatmaps (used to guide the diagnosis for labeling AD) overlaid on MRIs for two typical AD subjects, showing the activation distribution of a NeuroSymAD model.} 
  \label{fig:brainmri}  
\end{figure}

Fig. \ref{fig:brainmri} presents heatmaps overlaid on MRI brain scans in three different anatomical planes: sagittal, axial, and coronal, in a 3D ResNet model with NeuroSymbAD. The color intensity represents the magnitude of a specific activation or relevance score for being diagnosed with AD. The imaging patterns demonstrate that NeuroSymAD enables ResNet to focus successfully on critical regions in brain MRI scans, particularly highlighting areas around the ventricles, corpus callosum, and deep gray matter structures, where interpretation from our NeuroSymAD approach is closely aligned with current clinical findings.

\section{Conclusion}

We presented NeuroSymAD, a neuro-symbolic framework for AD diagnosis that combines neural networks for MRI perception with symbolic reasoning to mimic clinicians' diagnostic processes. By end-to-end optimization between perception and reasoning components, NeuroSymAD achieves superior performance while providing transparent diagnostic explanations, addressing critical limitations of current AI approaches and showing the potential of neuro-symbolic systems in healthcare applications where both accuracy and interpretability are essential.
\bibliographystyle{splncs04}
\bibliography{refer}

\end{document}